# Plug-and-play quantum devices with efficient fiber-quantum dot interface

*Woong Bae Jeon,[1] Jong Sung Moon,[1] Kyu-Young Kim,[1] Young-Ho Ko,[2] Christopher J. K. Richardson,[3] Edo Waks,[4,5] and Je-Hyung Kim[1]\**

[1]Department of Physics, Ulsan National Institute of Science and Technology (UNIST), Ulsan 44919, Republic of Korea

[2]Electronics and Telecommunications Research Institute (ETRI), Daejeon 34129, Republic of Korea

[3]Laboratory for Physical Sciences, University of Maryland, College Park, Maryland 20740, United States

[4]Department of Electrical and Computer Engineering and Institute for Research in Electronics and Applied Physics, University of Maryland, College Park, Maryland 20742, United States

[5]Joint Quantum Institute, University of Maryland and the National Institute of Standards and Technology, College Park, Maryland 20742, United States

**Incorporating solid-state quantum emitters into optical fiber networks enables the long-distance transmission of quantum information and the remote connection of distributed quantum nodes. However, interfacing quantum emitters with fiber optics encounters several challenges, including low coupling efficiency and stability. Here, we demonstrate a highly efficient fiber-interfacing photonic device that directly launches single photons from quantum dots into a standard FC/PC-connectorized single-mode fiber (SMF28). Optimally designed photonic structures based on hole gratings produce an ultra-narrow directional beam that matches the small numerical aperture of a single-mode fiber. A pick-and-place technique selectively integrates a single miniaturized device into the core of the fiber. Our approach realizes a plug-and-play single-photon device that does not require any optical alignment and thus guarantees long-term stability. The results thus represent a major step toward practical and reliable quantum lights across a fiber network.**

# 1. Introduction

Solid-state quantum dots (QDs) with a variety of exciton states provide important quantum building blocks such as single photons,[1-3] entangled photon pairs,[4,5] and spin memories[6] for quantum information processing. Recent advances in material growth, nanofabrication, and coherent control techniques for QDs have led to the realization of quantum communication,[5,7,8] quantum teleportation,[9,10] and quantum simulations.[3] Moreover, QDs can mediate the quantum interaction between spins and photons[11,12] or between photons and photons,[13] thus creating deterministic quantum gates[13,14] and photonic cluster states.[15] The problem of spectral inhomogeneity in solid-state quantum emitters can be addressed by temporal-to-spatial demultiplexing[16-18] and local frequency tuning[19,20] techniques. Multiple indistinguishable single photons and multiple resonant emitters from these techniques can scale quantum systems.

To implement the above quantum technologies in a practical way, interfacing the quantum emitters with low-loss optical platforms is essential. Well-developed fiber optics provide the ideal platform for establishing scalable and distributed quantum systems.[5,21] To efficiently and reliably transfer quantum information over fiber optic systems, many groups have demonstrated fiber-coupled quantum emitters. In the most straightforward way, the wafer, including QDs can be directly attached to an optical fiber.[22] This simple integration lacks coupling efficiency owing to the poor light extraction with non-Gaussian far-field patterns of the QD emission. Integrating QDs into micro/nanophotonic structures, such as micropillars,[23] monolithic lenses,[24-26] and bulls eye,[27,28] can significantly enhance the light extraction and produce Gaussian-like far-field patterns. However, the big mismatch in the numerical aperture (NA) between high-index nanophotonic structures and low-index single-mode fibers limits the coupling efficiency between them. Intermediate optical systems[29,30] or micro-structured optical fibers[31] can convert the size of optical modes to fit a single-mode fiber, but continuous adjustment of the optical alignment is inevitable to maintain coupling efficiency. Also, the above methods are based on chip-to-fiber integration, which wastes samples and limits the scalability. Ideally, adiabatic coupling between a tapered single-photon device and a tapered fiber can realize single device-to-fiber integration with potentially near-unit coupling efficiency.[32-34] However, this delicate tapered structure is sensitive to any movement, leading to long-term stability problems. Therefore, the efficient and reliable implementation of fiber-integrated quantum emitters remains a challenging goal.

Here, we demonstrate efficient plug-and-play single-photon sources based on telecom wavelength emitting QDs with optimal fiber-interfacing photonic structures. A thin-film planar resonator with a hole array produces an ultra-narrow vertical beam whose emission angle matches the small NA of a single-mode fiber. Compared to the well-known ring-based diffraction cavity, a so-called bulls eye cavity,[27,28,35,36] our hole-based diffraction cavity designs the structures in both radial and axial directions, thus optimizing the constructive and destructive interferences further than ring gratings. Using a hybrid integration technique,[37] we precisely integrate a single QD device into the core of a single-mode fiber. This integrated fiber–QD system shows distinctive advantages of high coupling efficiency, mechanical stability, and compatibility with a variety of platforms such as cryostats and electromechanical systems. Therefore, our approach realizes compact plug-and-play single-photon devices, as illustrated in **Figure 1a**.

## 2. Optimized design of the fiber-interfacing photonic device

Our photonic device includes hole-based circular Bragg gratings (*H*-CBG). The advantages of *H*-CBG over conventional ring-based circular Bragg gratings (*R*-CBGs) originate from several factors related to their hole-based structure. First, according to effective medium theory,[38,39] the hole gratings effectively have less contrast in the refractive index along a radial direction compared to ring gratings, for which abrupt high refractive contrast exists between air and the material (InP). This is important as a smooth refractive index gradient makes it easier to enlarge the spatial mode volumes. Moreover, *H*-CBGs optimize the structural parameters in both the radial and axial directions, whereas *R*-CBGs optimize the structure in only the radial direction. This additional degree of freedom possessed by *H*-CBGs enables further optimization of the constructive and destructive interferences, resulting in the highly enhanced vertical emission and suppressed higher-order diffraction.[40]

We designed the *H*-CBG using a finite-difference time-domain method and compared it with an *R*-CBG. For an *R*-CBG, the wavelength of the fundamental cavity mode mostly depends on the radial period ($\Lambda$) of the concentric rings that satisfy the second-order Bragg condition, $\Lambda = \lambda_{QD}/n_{eff}$ ($n_{eff}$ is the TE mode effective refractive index).[41] By changing the radius of the center disk (*c*) and the trench width between the rings (*w*), we optimized the structural parameters for achieving the maximum collection efficiency within a half-angle of 7°, which corresponds to the NA of a standard single-mode fiber (SMF28). The optimized values of the structural parameters for the *R*-CBG were $\Lambda$ = 497 nm, *c* = 546 nm, and *w* = 149 nm (Figure S1). The thickness (*T*) of the structure was fixed at 280 nm in order to match the QD sample. To imitate QDs, we inserted two orthogonally polarized in-plane dipoles at the center of a structure. In the case of the *H*-CBGs, we similarly simulated a structure with radial parameters of the radial period between holes ($\Lambda$), the radius of the center disk ($c$). The hole size ($h$) was considered instead of the trench width ($w$) in the *R*-CBG. We note that the *H*-CBG structure has an additional parameter of the axial period between holes ($a$). The optimized structures had the structural parameters of $\Lambda = 467$ nm, $c = 618$ nm, $h = 135$ nm, and $a = 210$ nm (Figure S1). Figure 1b and c exhibits the structural design with a superimposed in-plane near field $|E|^2$ profile of the *R*- and *H*-CBGs, respectively. The in-plane spatial mode of the *H*-CBG is distributed over the entire device, which is in contrast to the tightly confined mode in the central disk of the *R*-CBG. The larger *H*-CBG mode size leads to more directional emission than the case of the *R*-CBG. This is shown in the vertical cross-sectional field ($|E|^2$) profiles for the *R*- and *H*-CBGs in Figure 1b and c. The difference in the emission angle between the two structures is clearly visible from the far-field patterns plotted against the angle (Figure 1d). For the *H*-CBG, the most emission lies within the NA = 0.12 of a single-mode fiber (SMF28). On the other hand, although the *R*-CBG has a Gaussian far-field pattern, a considerable coupling loss is expected when coupled to a single-mode fiber. Figure 1e compares the cross-sectional far-field profiles of Figure 1d, and it shows that the *H*-CBG has a suppressed high-order diffraction beam. In contrast, the *R*-CBG produces noticeable emissions into higher-order diffraction angles greater than 30°. Figure 1f quantitatively compares the collection efficiencies of the *R*- and *H*-CBGs at the center wavelength of the cavity mode, calculated as the fraction of the total upward emitted light that passes within the various NAs. As expected, the *H*-CBG maintains high collection efficiencies even at small NAs. In the case of the *R*-CBG, only 24% of the emitted light can be collected with 0.12 NA, whereas the *H*-CBG can collect as much as 63%.

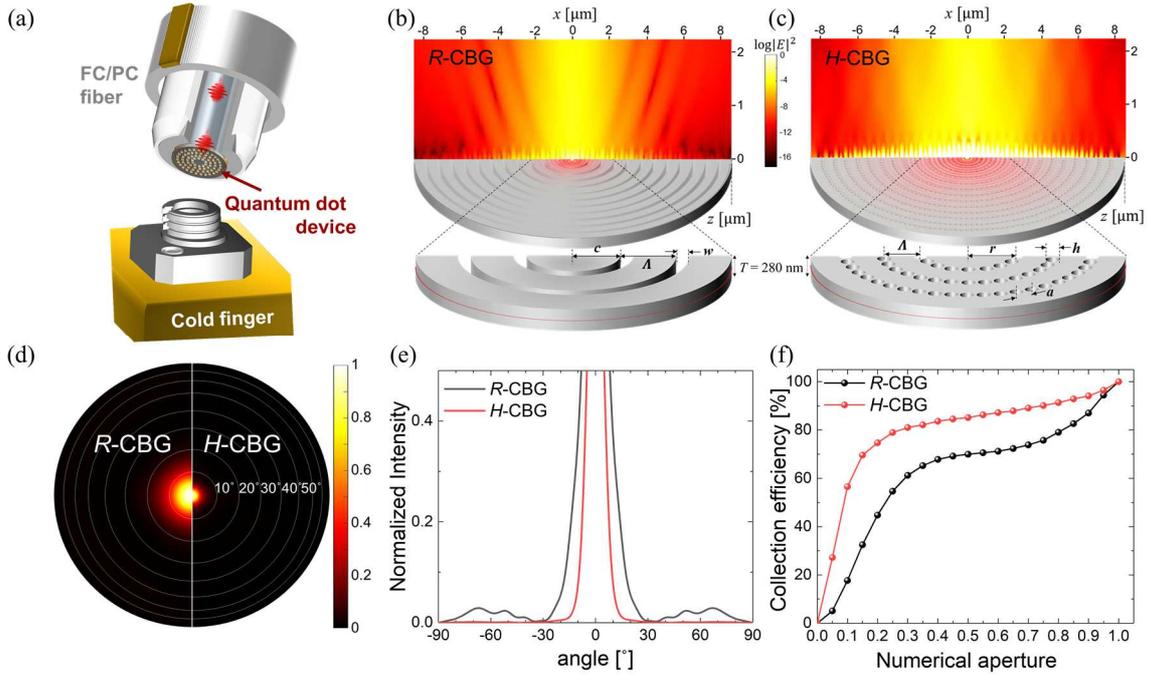

**Figure 1.** Ultra-narrow vertical beam in an H-CBG. a) Schematic image of an H-CBG device integrated into a single-mode fiber. The H-CBG device includes InAs/InP QDs. The integrated fiber–QD device can be stably mounted on a cold fiber using an FC/PC mating sleeve for low-temperature operation. b) and c) illustrate schematics and simulation results for an R-CBG and an H-CBG, respectively. The vertical cross-sectional field profiles show the normalized $|E|^2$ for each device on a logarithmic scale. The in-plane spatial field ($|E|^2$) profiles for the fundamental mode are superimposed on the top of each structural schematic. d) Polar plots of far-field mode profiles. The left-hand (right-hand) semi-circle corresponds to an *R*-CBG (*H*-CBG). e) Cross-sectional far-field mode profiles as a function of the angle. f) Calculated collection efficiency as a function of the NA.

## 2. Device fabrication and optical characterization

The sample contained self-assembled InAs QDs with a density of approximately 10 µm$^{-2}$ located in the middle of a 280 nm-thick InP membrane on a 2 µm-thick AlInAs sacrificial layer. The emissions of the QD ensemble span a broad spectral range from 1200 nm to 1550 nm, covering the telecom O-and C-bands. In this study, the short-wavelength region around 1250 nm, where the density of the QDs has a relatively low density, was designated as the center of the cavity mode to obtain single QD emissions with low background fluorescence. The single-photon devices were fabricated on an InAs/InP QD wafer using electron-beam lithography followed by dry and wet etching processes. **Figure 2a** shows a scanning electron microscopy (SEM) image of the fabricated air-suspended *H*-CBG. In the simulation, the fabricated *H*-CBG has a Q-factor of 4200, leading to a Purcell factor over 100 near the cavity mode (Figure 2b). This is considerably larger than the Q-factor (230) and Purcell factor (20) of the *R*-CBG (Figure S1). Although the high Purcell enhancement occurs within a small spectral window within the FWHM of 0.38 nm at the resonant wavelength, the enhancement in the collection efficiency takes place over a broad spectral range, as shown in Figure 2b. This is advantageous for spectral coupling with solid-state quantum emitters.

For low-temperature micro-photoluminescence (PL) study of the *H*-CBGs, we excited the sample using a 785 nm laser and collected the emissions using a microscope objective lens (0.7 NA) in free space. Figure 2c shows the PL spectra of a cavity mode and coupled QDs at high (500 nW) and low (10 nW) excitation powers. At the low excitation power, the ground state emissions of the QDs are visible. At the high excitation power, the multi-excitonic states of the QDs and their continuous hybridized states with a wetting layer efficiently feed the cavity modes so that the cavity emission dominates over the single QD emissions. The fabricated cavity has a Q value of about 700. Unlike the simulated mode spectrum shown in Figure 2b, the measured mode spectrum shows multiple peaks near the simulated fundamental cavity mode. These multiple peaks originate from the structural asymmetry of the fabricated device (Figure S2).

The single QDs exhibit significantly increased PL intensities near the cavity mode. To prove that the bright single QD peak corresponds to single-photon emission, we performed a second-order correlation ($g^{(2)}(\tau)$) measurement using a Hanbury-Brown and Twiss setup. The measured histogram shows an antibunching signal with $g^{(2)}(0) = 0.16 \pm 0.02$, indicating a single-photon emission. The non-zero $g^{(2)}(0)$ is mostly due to the spectrally superimposed broad cavity emission. To quantitatively measure the brightness of single photons from the *H*-CBG devices, we estimated the collection efficiency at the first lens (0.7 NA) by carefully calibrating the system efficiency (0.64 ± 0.04%), including the detector efficiency (20%) (Figure S3). To correct the multi-photon probability, we also multiplied the collection efficiency by the multi-photon correction term $\sqrt{1 - g^{(2)}(0)} = 0.778$.[42] The QD was excited with a 40 MHz pulsed laser, and the photon counts were monitored at the saturation power ($P_{sat}$). Since the photon counts exceeded the saturation level of the InGaAs single-photon detectors, we inserted a neutral density filter (8.7 dB) before the detector and took this into account during the calibration. Considering the above factors, we calculated the collection efficiency of two different QDs (QD A and QD B) in *H*-CBGs with different amounts of spectral detuning (0.1 nm and 4 nm, respectively) and compared the values with that of bulk QDs (Figure S4). In Figure 2d, the bulk QDs show an expected low collection efficiency of less than 1%, whereas QD A and QD B in the *H*-CBG exhibit a much higher collection efficiency of 24.3% and 30.1%, respectively. The results represent that the large enhancement in the collection efficiency occurs over a wide spectral range, as expected from the simulation in Figure 2b. We attribute the discrepancy between the simulated and measured collection efficiencies at 0.7 NA to the spatial mismatch between the QDs and the cavity mode and to the imperfection in the fabricated device (Figure S5). Also, it should be noted that the simulation only considered the upward emission from the device. Therefore, to achieve the simulated collection efficiency, a good bottom reflector is required to fully reflect the downward emission.

Besides increasing the brightness, the cavity engineers the recombination dynamics of coupled QDs. Figure 2e compares the decay curves of cavity-coupled QDs A and B with the typical decay curve of a bulk QD. The bulk QDs had an average decay time of around 2 ns. In the case of QD A, it is closely resonant with the cavity mode and shows a fast decay time of 0.66 ns, while QD B, with the large spectral detuning of 4 nm, shows a slow decay time of 3.07 ns. Figure 2f displays a statistical distribution of the decay times of QDs in different *H*-CBGs as a function of the spectral detuning. A Purcell enhancement of up to four near the cavity mode can be observed in the fabricated *H*-CBGs.

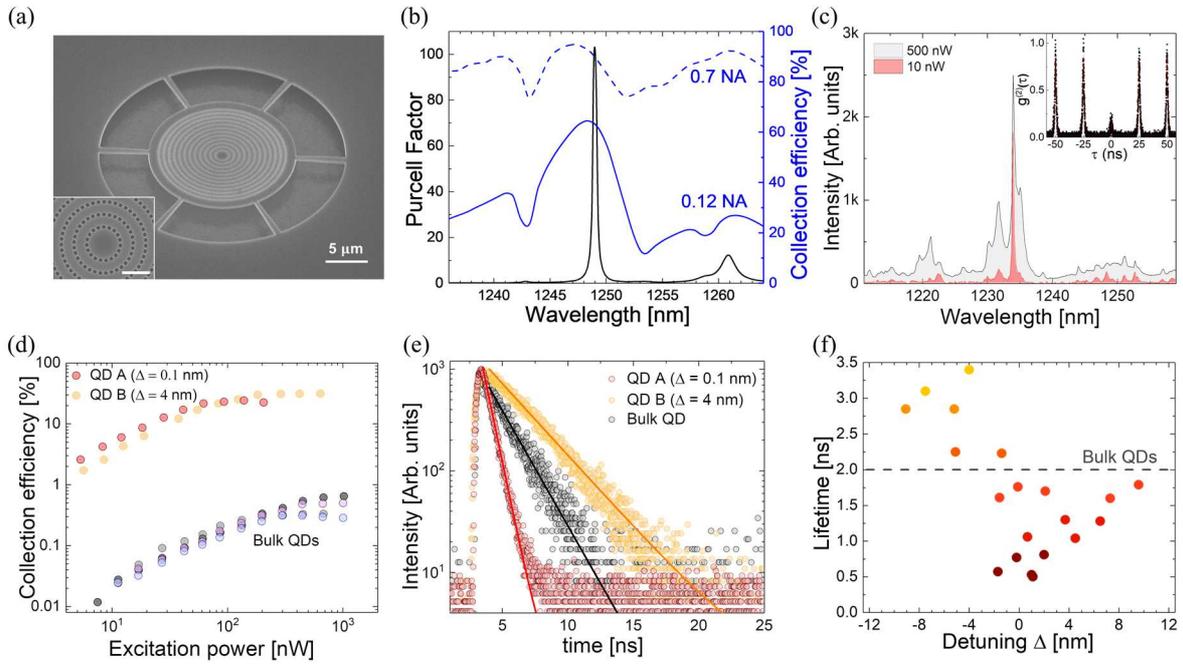

**Figure 2.** Optical characteristics of the *H*-CBG device. a) SEM image of the fabricated *H*-CBG. The inset shows a close-up view of the center of the device. The scale bar in the inset is 1 $\mu$m. b) Calculated Purcell factor (black line) and collection efficiency for an NA of 0.7 (dashed blue line) and 0.12 (solid blue line), corresponding to the objective lens and the single-mode optical fiber respectively. c) PL spectrum of an *H*-CBG at 500 nW (gray curve) and 10 nW (red curve). The inset shows an antibunching curve for the single QD emission, showing $g^{(2)}(0) = 0.16 \pm 0.02$. d) Collection efficiency of QD A (red circle) and QD B (orange circle) in the *H*-CBG with different spectral detuning of 0.1 nm and 4 nm, respectively. Collection efficiencies of four different bulk QDs (gray, dark gray, blue, purple) are plotted together for comparison. e) Decay curves for QDs A (red circle) and B (orange circle) in the *H*-CBG together with that of a bulk QD (black circle). The solid lines are fitted single-exponential curves with decay times of 0.66 ns (red), 3.07 ns (orange), and 2 ns (black). (f) Distribution of decay times of various cavity-coupled QDs as a function of the spectral detuning (Δ) between the cavity and the QDs. The dashed black line corresponds to the average decay time of the bulk QDs.

## 3. Integration of a single H-CBG device into a single-mode fiber

An important figure-of-merit for fiber-coupled quantum emitters is the coupling efficiency of single photons over a single-mode fiber. To demonstrate the direct integration of a single QD device into a fiber, we employed a pick-and-place technique based on a polydimethylsiloxane (PDMS) stamp. The fabricated *H*-CBG devices were air-suspended by thin tethers in the QD wafer. As described in **Figure 3a and b**, the transparent PDMS stamp on a glass slide enables us to transfer a single *H*-CBG from the QD wafer to the core of a fiber while monitoring in real-time under an optical microscope. To align the central axis of the device and the core of the fiber, a red (633 nm) laser beam was sent from the opposite end of the fiber. Figure 3b shows the optical image of the end facet of the single-mode fiber after the integration of the single *H*-CBG device. The *H*-CBG device is firmly attached to the fiber by van der Waals force and, as shown in the inset in Figure 3b, is well-centered at the fiber core. The integrated fiber–QD device was then covered by a 50 nm-thick Au film that serves as a partial back reflector and thermal conductor for the low-temperature measurement.

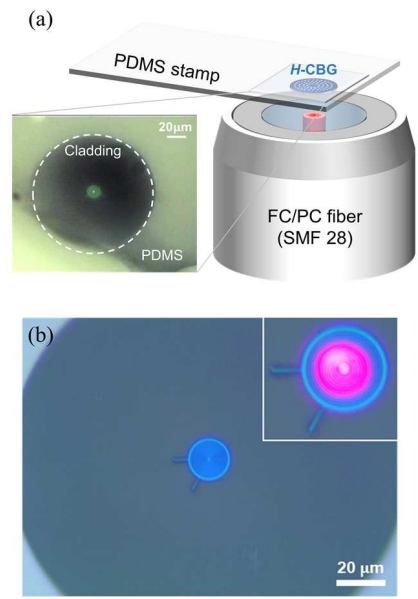

**Figure 3.** Integration of a single *H*-CBG device into an FC/PC fiber. a) Schematic of the integration process between a single *H*-CBG and a single-mode fiber (SMF28) using a PDMS stamp. The inset is an optical microscopy image (×50) showing that the *H*-CBG is aligned with the fiber core during the transfer process. b) Optical microscopy image of the *H*-CBG transferred onto the end facet of the fiber. The inset shows a close-up image of the fiber core illuminated by a 633 nm laser beam sent from the opposite end of the fiber.

## 4. Bright and stable single photons in an all-fiber coupled system

**Figure 4a** displays a photographic image of the fiber-QD devices mounted on a cold finger for low-temperature measurements. The use of a FC/PC-connectorized fiber enables the fiber–QD device to be mounted on a cold finger using conventional FC/PC mating sleeves. After mounting the integrated fiber–QD device in the cryostat, the fibers both inside and outside the cryostat were connected using a fiber feedthrough installed in the cryostat lid. We sent a 785 nm pumping laser and collected the QD emission through the fiber using a 90:10 fiber beam splitter. To measure single photons, we spectrally filtered out a single QD emission using a fiber-based tunable filter and sent single photons to a fiber-based 50:50 beam splitter followed by fiber-coupled InGaAs single-photon detectors. Figure 4b describes a schematic of an all-fiber coupled single-photon system from the source to the detector. Figure 4c compares the PL spectrum of bulk QDs measured in free space (top) with that of the *H*-CBG directly coupled to a single-mode fiber (bottom). The emissions close to 970 nm correspond to a wetting layer. In free space, the bulk QDs exhibits very weak emissions from 1200 to 1550 nm even with a high NA = 0.7 objective lens. For the fiber–QD device, the PL spectrum clearly shows that the *H*-CBG device efficiently out-couples the QD emissions through the single-mode fiber (0.12 NA). For the spectrally coupled QDs with the cavity mode, their emissions increase as large as the wetting layer emission, which was two orders of magnitude higher than that of bulk QDs in the free space setup.

For the brightest single QD peak in the fiber–QD device, we measured a $g^{(2)}(\tau)$ curve and brightness. The observed $g^{(2)}(0) = 0.16 \pm 0.02$ from the fiber–QD device at 0.1 $P_{\text{sat}}$ is similar to the value in the free space (Figure 4d). To estimate the brightness, we excited the QD at a low repetition rate of 2.5 MHz, which prevents saturation effects in an InGaAs detector and observed a single photon detection rate of $10.63 \pm 0.11$ kHz at $P_{\text{sat}}$. To deduct the QD-to-fiber coupling efficiency, we separately measured the system efficiency ($4.6 \pm 0.3\%$), including the detector efficiency (20%) (Figure S3), and then calibrated the multi-photon correction term $\sqrt{1 - g^{(2)}(0)} = 0.872$ at the saturation power. This results in the QD-to-fiber coupling efficiency of $8.1 \pm 0.5\%$. If we consider a typical 80 MHz pulsed excitation, this value corresponds to a single-photon rate of 6.5 MHz at the end of the first fiber. In the simulation, the mode coupling efficiency between the *H*-CBG and the single-mode fiber was around 53% (Figure S6). Although the achieved coupling efficiency is lower than the simulated value, the measured single-photon coupling efficiency is a record high for fiber-integrated QD devices at telecom wavelengths.[26,33,43] The non-ideal position of the QD in the *H*-CBG device, the spatial misalignment between the *H*-CBG device and the fiber core, and imperfect fabrications, such as an angled etch profile, could be attributed to the lower coupling efficiency in the experiment (Figure S5 and S6). In addition, InAs/InP QDs might have non-unit internal efficiency even at low temperatures.[44] The measured value, therefore, gives a lower bound on the possible coupling efficiency.

Finally, the single-photon count rates at $P_{sat}$ of the fiber-integrated single-photon device are plotted against time in Figure 4e. We observed the intensity fluctuations within 1%, mostly originating from the fluctuation of the laser power. Such stable count rates are maintained even after multiple thermal cycles. Since any optical alignments are not necessary for an all-fiber coupled single-photon system, the system ensures long-term stability.

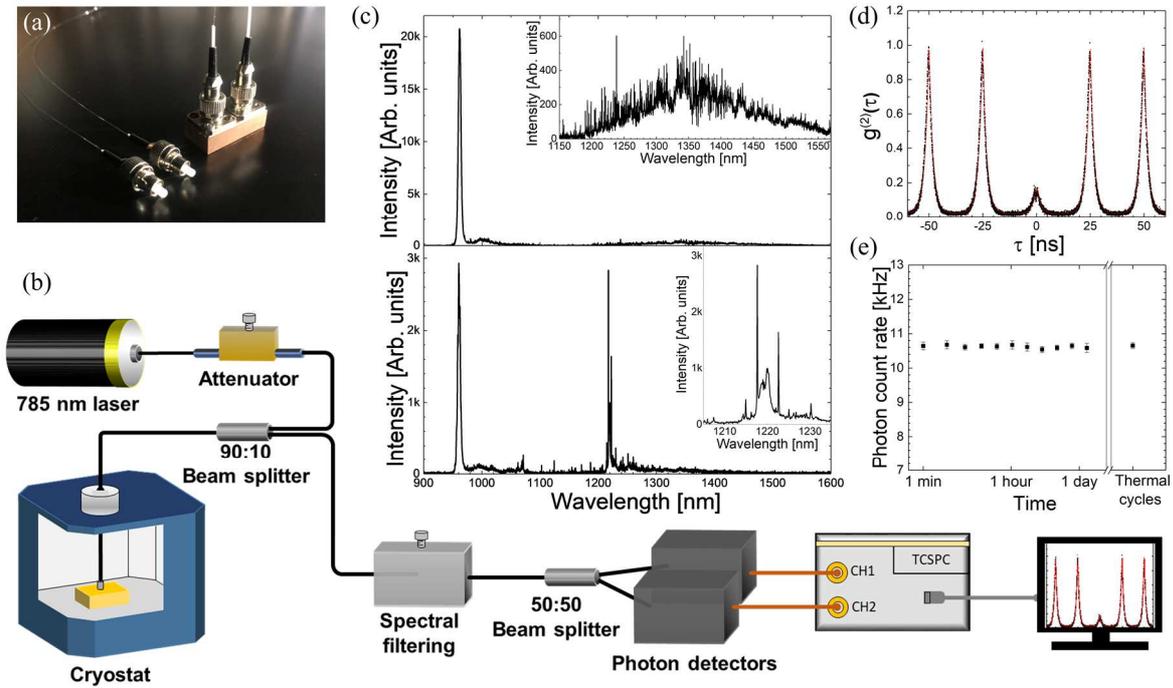

**Figure 4.** All-fiber coupled single-photon system. a) Photograph of the fiber–QD device installed on a cold finger. b) Schematic of an all-fiber coupled optical setup from the source to the detectors. c) Comparison of the PL spectra between the bulk QD sample measured in free space (0.7 NA) (top) and the single $H$-CBG device integrated with a single-mode fiber (0.12 NA) (bottom). The peak near 970 nm corresponds to a wetting layer. The insets are close-up views of the QD emissions of each spectrum. d) Autocorrelation curve ($g^{(2)}(\tau)$) for the single QD emission from the fiber–QD device. e) Intensity plot of the single-photon count rate of the fiber–QD device against time and thermal cycles.

## 5 Conclusion

In this work, we have successfully demonstrated efficient and reliable transmission of single photons from single QDs into an all-fiber coupled system. A hole-based diffraction cavity generates a narrow, vertical beam that matches the small NA of a standard single-mode fiber at telecom wavelengths. This highly efficient photonic interface enables the simple and robust integration of a single QD device into an FC/PC-connectorized fiber. This provides practical all-fiber platforms from the sources to the detectors for optical quantum information processing. By introducing position control techniques for the QDs[27] and by using more precise pick-and-place techniques,[45] a coupling efficiency up to the theoretical limit could be achieved. The inclusion of an optimized back reflector can further increase the coupling efficiency.[36] The configuration of the hole arrays in the *H*-CBG could be optimized further by adding more variations in the hole size and positions instead of fixed hole sizes and periodic positions used in this study. Machine learning techniques will be adequate for optimizing such variations.[46] With the rapid development of quantum photonic technologies, the demand for practical quantum light sources is increasing in various fields, including quantum imaging, quantum diagnostics, and quantum simulation using photons. Our approach could provide practical and reliable plug-and-play quantum light sources and be extended to scalable quantum systems by integrating several pre-characterized quantum emitters into multiple fibers, such as V-groove fiber arrays. Moreover, integrated quantum emitters with spin states could implement quantum memories and repeaters in fiber quantum networks. Therefore, our approach paves the way for distributing quantum lights and connecting quantum nodes on commercialized fiber platforms.

**Acknowledgements**

The authors would like to acknowledge support from the National Research Foundation of Korea grant funded by the Korea government (MSIT) (NRF-2020M3H3A1098869, NRF-2022R1A2C2003176) and Institute of Information and Communications Technology Planning and Evaluation (IITP) grant funded by the Korea government (MSIT) (No.2019-0-00434), the ITRC (Information Technology National Research Center) support program (IITP 2021-2020-0-01606) supervised by the IITP, and the Air Force Office of Scientific Research (Grant No.FA23862014072).

**Conflict of Interest**

The authors declare no conflict of interest.

# Supporting Information

## 1. Numerical simulation of ring- and hole-based cavities

We optimized the designs of ring-based circular Bragg gratings (*R*-CBGs) and hole-based CBGs (*H*-CBGs) using finite-difference time-domain simulation. First, we designed the *R*-CBG based on a standard circular Bragg grating model[1,2] with a radial period (*Λ*) defined as $\Lambda = \lambda_{QD}/n_{eff}$. The radius of the center disk (*c*) is chosen to 1.1*Λ*. Then we sweep the trench width (*w*) between rings and the radius of the center disk (*c*) of the *R*-CBGs to yield the highest collection efficiency within a numerical aperture (NA) of 0.12. The *H*-CBG was optimized in a similar way with parameters of the center disk (*c*) and hole size (*h*), but the *H*-CBG has one more degree of freedom in the axial direction; axial period (*a*) between neighbored holes. A schematic image with the designed parameters is depicted in Figure S1a. From these parameters, we could find an optimized relationship between parameters *a* and *h* for the maximum collection efficiency within an NA = 0.12. Such relation is shown in Figure S1b. The size of a center disk (*c*) and radial period (*Λ*) determined the wavelength of the cavity mode. Figure S1c and S1d show the cavity mode spectrum of the *R*-and *H*-CBGs with different *Λ*. In the optimized conditions, the *R(H)*-CBGs had the structural parameters of *Λ* = 497(467) nm, *c* = 546(618) nm, and *w(h)* = 149(135) nm. The axial period of holes for the *H*-CBG was *a* = 210 nm.

Figure S1e and S1f plotted the Purcell factors and collection efficiencies of the *R*-and *H*-CBGs. As expected from the narrow cavity spectrum of the *H*-CBG, the *H*-CBG has a high Purcell factor up to 100, relatively larger than that of the *R*-CBG around 20. In terms of collection efficiency, as both structures generate Gaussian far-field patterns, they have high collection efficiencies of over 90% within a half-angle of 44.43°, corresponding to the NA = 0.7. However, if we consider the collection efficiency within a small NA = 0.12, the value drops below 25% in the *R*-CBG, whereas the high collection efficiency is maintained up to 63% in the *H*-CBG even at a small NA = 0.12.

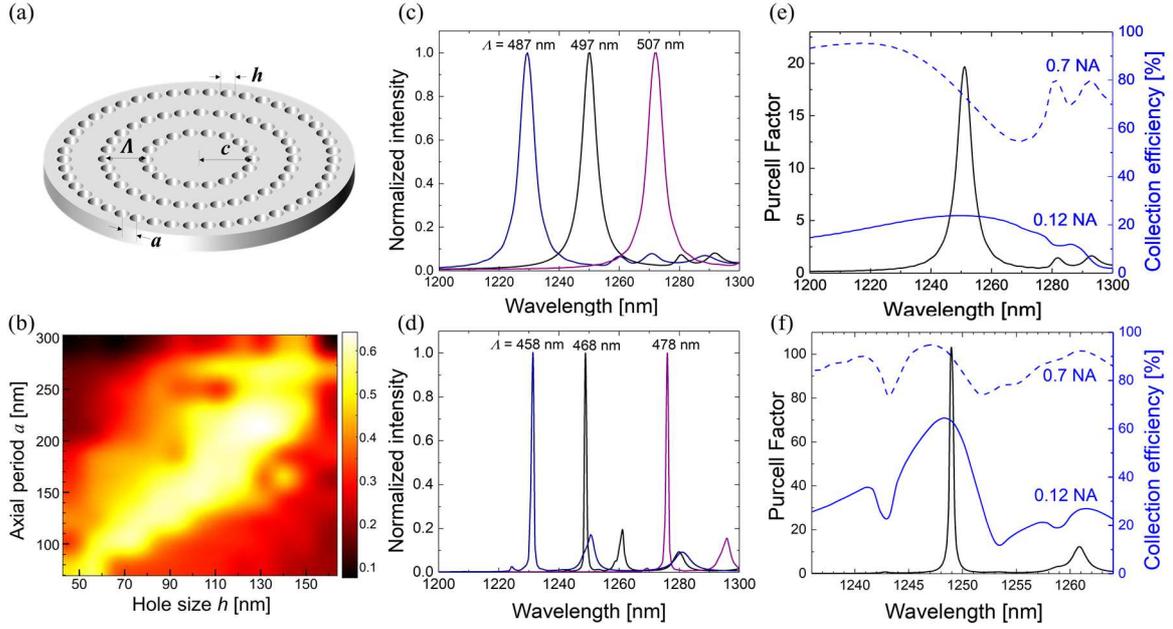

**Figure S1**. a) Schematic image of an *H*-CBG having structural parameters of the radius of the center disk (*c*) and hole size (*h*), radial period (*Λ*), and axial period (*a*) between holes. b) 2D plot of calculated collection efficiency as a function of parameters *a* and *h*. c) and d) show cavity mode spectra with the different radial period (*Λ*) of the *R*-CBG and the *H*-CBG, respectively. e) and f) show Purcell factor (black line) and collection efficiency of the *R*-CBG and the *H*-CBG in the 0.12 NA (solid blue line) and 0.7 NA (dashed blue line), respectively.

## 2. Asymmetry in hole patterns

To figure out the origin of multiple peaks of the cavity spectrum in the fabricated *H*-CBG, we simulated the *H*-CBG structures with elongated holes that frequently occur during the fabrication process. Figure S2a and S2b compares the simulated cavity modes with ideal circular holes to that with elongated holes having a ratio of $h_x : h_y = 1:1.05$. The resulting spectrum with such elongated holes is very similar to the measured cavity spectrum (Figure S2c).

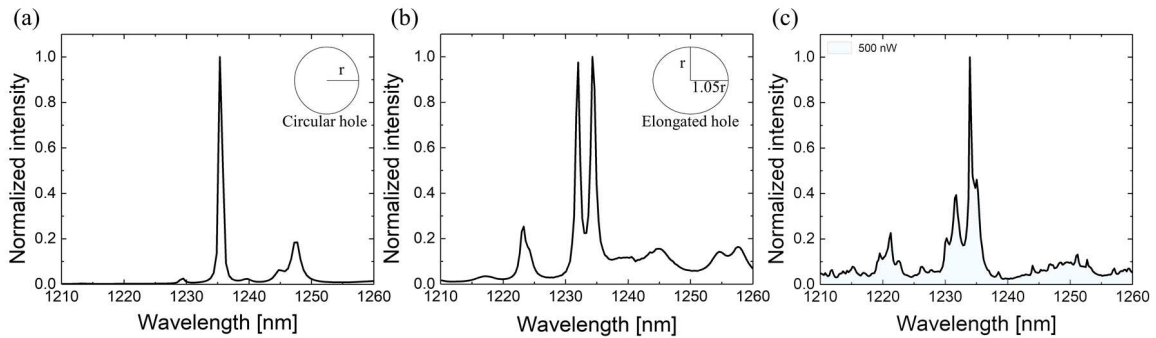

**Figure S2.** Simulated mode spectra of the *H*-CBG with ideal circular holes a) and with elongated holes b). The ellipse hole has an eccentricity of 0.305. c) Measured cavity mode spectrum of the fabricated *H*-CBG.

## 3. Collection efficiencies in free space and fiber-based systems

To estimate the collection efficiency of single photons at the first lens in free space and the coupling efficiency of single photons at the first fiber in an all-fiber coupled system, we measured the transmission efficiency of the measurement systems in free space and fiber optics. The schematics of each setup are shown in Figure S3a and S3b. The tables summarize the efficiencies of individual parts of each setup, including the InGaAs single-photon detector (20% from the product specification). Overall, the free-space system has a system efficiency of around 0.64%, and the fiber optic system has a system efficiency of around 4.6%.

To measure the collection efficiency, we excited a single quantum dot (QD) with a 785 nm pulsed laser and measured the single-photon detection rate in each setup. To avoid the saturation effect, we inserted a neutral density filter (8.7 dB) before the detector in the free space system and took this into account for the photon counts. For the all-fiber coupled system, instead of a neutral density filter, we lowered the excitation rate up to 2.5 MHz to eliminate the count rates by the detectors, we separately measured the dark counts and after pulse signals and subtracted these values from the measured count rate. We also considered the multi-photon probability from the $g^{(2)}(0)$ value and obtained corrected detector count rates from each system. Finally, we deducted the single-photon collection efficiency of 24.3% and 30.1% for two QDs A and B and the single-photon fiber coupling efficiency of 8.1% for the fiber–QD device.

(a)

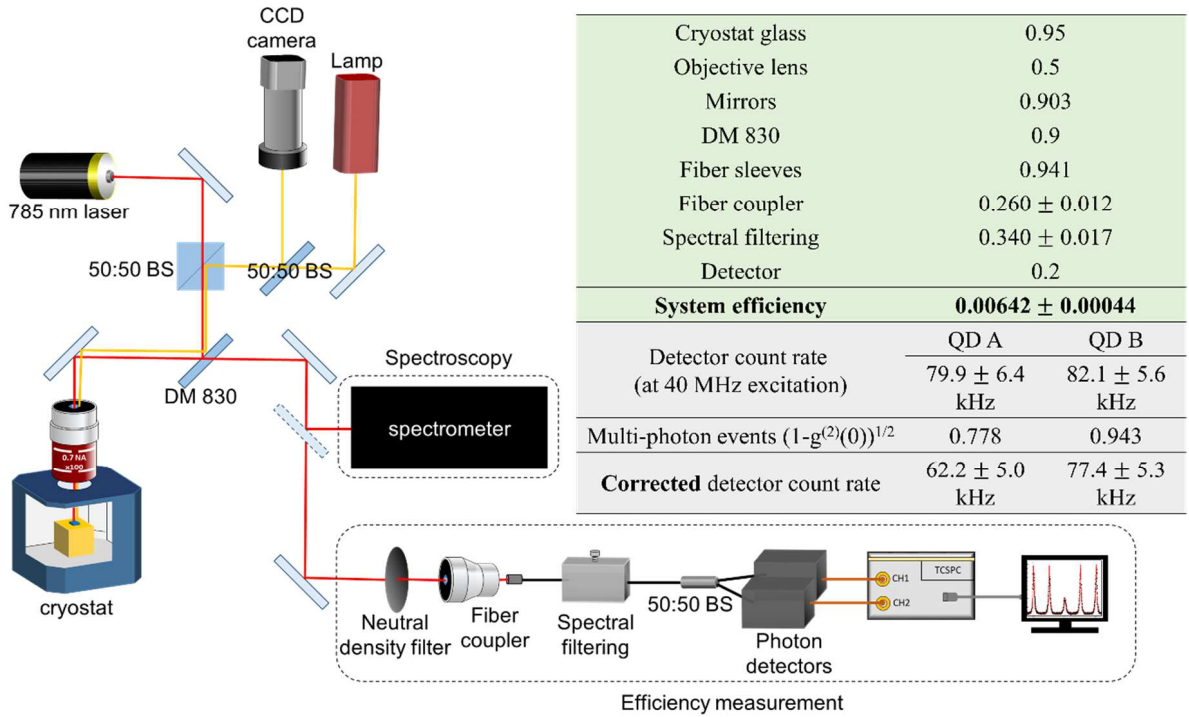

(b)

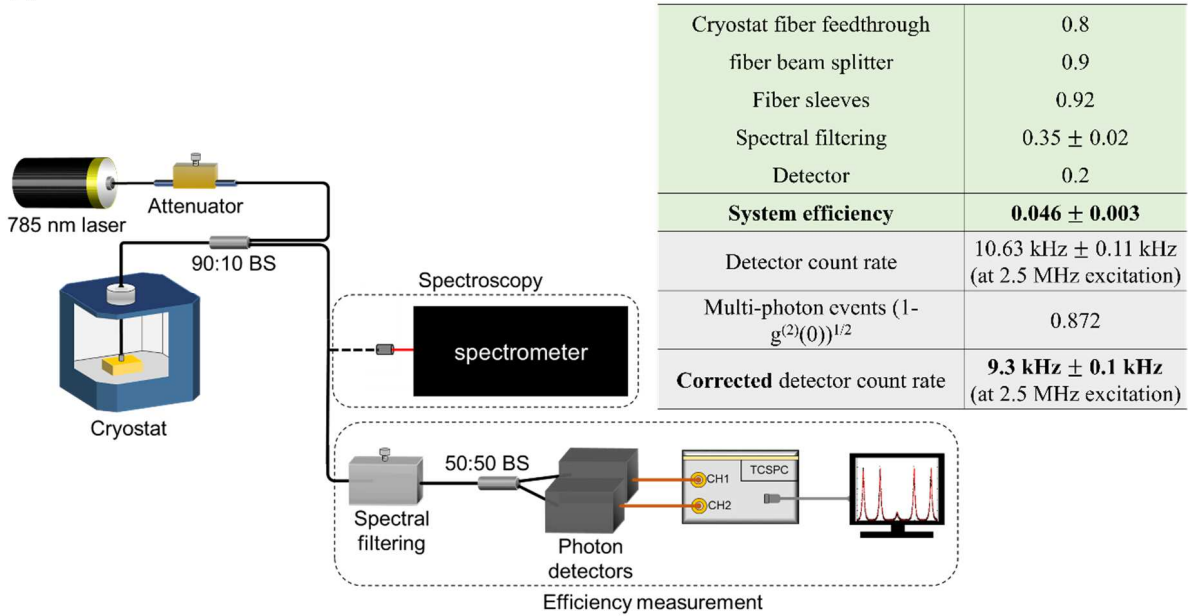

**Figure S3.** Schematic and efficiency tables for the free space a) and all-fiber coupled system b).

## 4. PL spectrum of the spectrally detuned QD B in the *H*-CBG

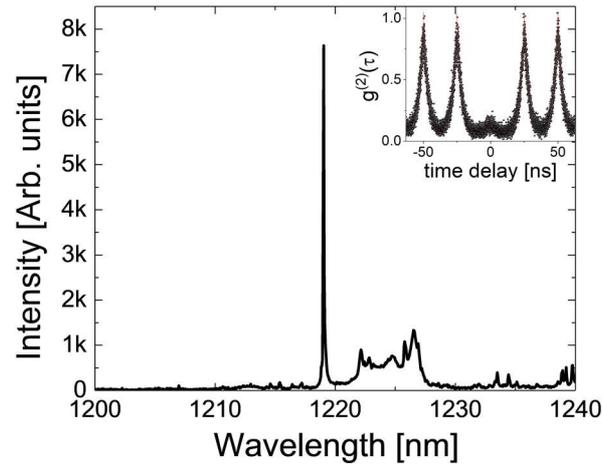

**Figure S4.** PL spectrum of spectrally detuned ($\Delta$ = 4 nm) QD B from the center of the cavity mode of the *H*-CBG. The inset shows an antibunching curve of the single quantum dot emission, showing $g^{(2)}(0) = 0.11 \pm 0.01$.

## 5. Simulation of spatial mismatch of QDs and fabrication imperfection

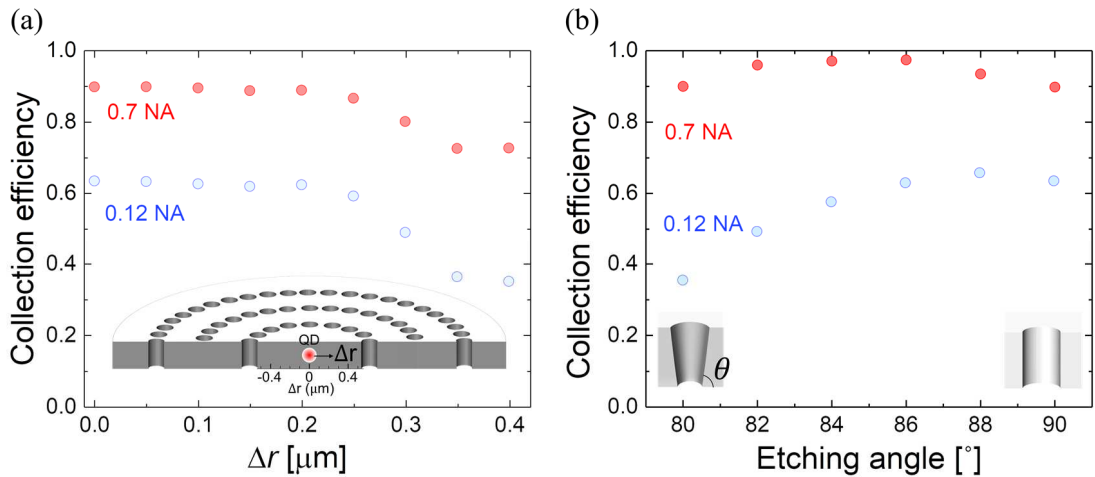

**Figure S5.** a) Collection efficiencies against the spatial misalignment ($\Delta r$) of the QD from the center of the *H*-CBG at the NA = 0.7 (red circles) and NA = 0.12 (blue circles). b) Collection efficiencies according to various etching angles of the hole pattern at the NA = 0.7 (red circles) and NA = 0.12 (blue circles).

## 6. Simulation of spatial mismatch between the *H*-CBG and a single-mode fiber

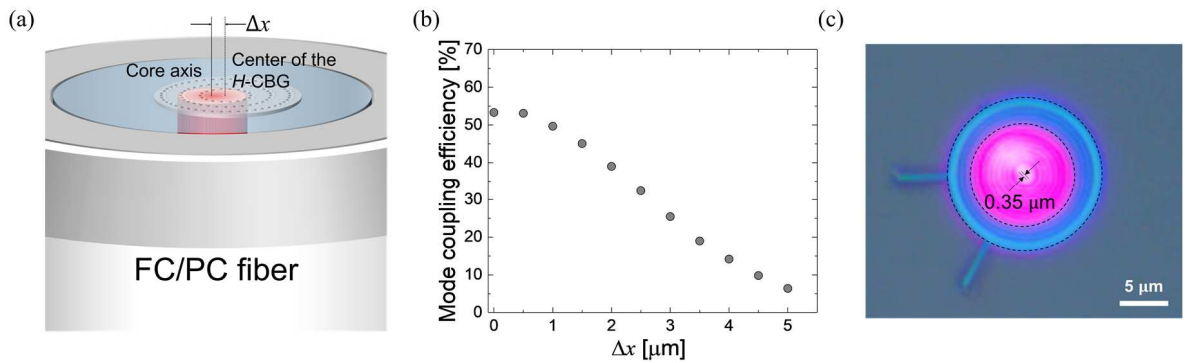

**Figure S6.** a) Schematic image of the spatial mismatch ($\Delta x$) between the *H*-CBG and the FC/PC fiber core. b) Mode coupling efficiency between the *H*-CBG and the single-mode fiber as a function of $\Delta x$. c) Optical image of the fiber-integrated *H*-CBG. As an alignment guide, a 633 nm laser was coupled to the opposite end of the fiber.